\documentclass[12pt,tightenlines,floats,aps,amsmath,amssymb,nofootinbib,superscriptaddress,showpacs]{revtex4}

\usepackage{amsmath,amsthm,amssymb,amsfonts}
\usepackage{color}

\begin{document}

\title{The combinatorics of the $SU(2)$ black hole entropy in loop quantum gravity}

\author{Iv\'an  \surname{Agull\'o}}
\email[]{Ivan.Agullo@uv.es} \affiliation{ {\footnotesize Physics Department, University of
Wisconsin-Milwaukee, P.O.Box 413, Milwaukee, WI 53201 USA}} \affiliation{Departamento de
F\'{\i}sica Te\'orica and IFIC, Centro Mixto Universidad de
Valencia-CSIC. Facultad de F\'{\i}sica, Universidad de Valencia,
Burjassot-46100, Valencia, Spain}

\author{J. Fernando \surname{Barbero G.}}
\email[]{fbarbero@iem.cfmac.csic.es} \affiliation{Instituto de
Estructura de la Materia, CSIC, Serrano 123, 28006 Madrid, Spain}

\author{Enrique F.  \surname{Borja}}
\email[]{Enrique.Fernandez@uv.es} \affiliation{Departamento de
F\'{\i}sica Te\'orica and IFIC, Centro Mixto Universidad de
Valencia-CSIC. Facultad de F\'{\i}sica, Universidad de Valencia,
Burjassot-46100, Valencia, Spain}

\author{Jacobo  \surname{D\'{\i}az-Polo}}
\email[]{Jacobo.Diaz@uv.es}\affiliation{Departamento de Astronom\'{\i}a y
Astrof\'{\i}sica, Universidad de Valencia, Burjassot-46100,
Valencia, Spain}

\author{Eduardo J. \surname{S. Villase\~nor}}
\email[]{ejsanche@math.uc3m.es} \affiliation{Instituto Gregorio Mill\'an, Grupo de Modelizaci\'on
y Simulaci\'on Num\'erica, Universidad Carlos III de Madrid, Avda.
de la Universidad 30, 28911 Legan\'es, Spain} \affiliation{Instituto
de Estructura de la Materia, CSIC, Serrano 123, 28006 Madrid, Spain}

\date{Jul 22, 2009}

\begin{abstract}
We use the combinatorial and number-theoretical methods developed in previous work by the authors to study black hole entropy in the new proposal put forward by Engle, Noui and P\'erez. Specifically we give the generating functions relevant for the computation of the entropy and use them to derive its asymptotic behavior including the value of the Immirzi parameter and the coefficient of the logarithmic correction.
\end{abstract}


\maketitle


In this brief note we want to study some of the physical consequences that follow  from the black hole entropy definition proposed, in the context of loop quantum gravity,  by Engle, Noui and P\'erez in \cite{ENP}. The main reason to do this is to check wether this new definition satisfies the obvious physical requirement of reproducing the Bekenstein-Hawking formula for large black holes.  Without going into the details of the theoretical foundations of this new proposal, this analysis can be seen as a straightforward consistency check. We also want to obtain corrections to this formula that can be eventually compared with equivalent results found in different approaches \cite{ABCK,ABK,DL,GM}. An additional reason to perform this study is to show the power of the combinatorial methods developed by the authors in \cite{prlnos,EF,EF1,val}.

The problem of interest can be enunciated in the following way \cite{ENP}. Given a value of the black hole area $a_H=4\pi\gamma\ell^2_P \kappa$ (where $\kappa\in \mathbb{N}$ is the level of the $SU(2)$ Chern-Simons theory on the horizon,\footnote{For an earlier treatment, based on different considerations, of the $SU(2)$ Chern-Simons theory in this framework see \cite{KM}.} $\ell_P$ denotes the Planck length, and $\gamma$ the Immirzi parameter),  we have to determine the number of states labeled by spins $j_1,\ldots, j_n$ satisfying an inequality of the type
\begin{eqnarray}
\label{ineq}
a_H-\epsilon\leq 8\pi\gamma \ell^2_P\sum_{p=1}^{n}\sqrt{j_p(j_p+1)}\leq a_H+\epsilon
\end{eqnarray}
or alternatively, following the prescription given in \cite{DL},
\begin{eqnarray}
\label{ineqDL}
8\pi\gamma \ell^2_P\sum_{p=1}^{n}\sqrt{j_p(j_p+1)}\leq a_H\,.
\end{eqnarray}
Each of these lists gives a contribution to the entropy equal to the  dimension of the Hilbert space $\mathcal{H}^{\rm CS}(j_1,\ldots,j_n)$ of the Chern-Simons theory associated with the fixed choice
of spins $j_p$ at each puncture $p$ of the horizon. When the Immirzi parameter satisfies $|\gamma|\leq \sqrt{3}$ the space $\mathcal{H}^{\rm CS}(j_1,\ldots,j_n)$ coincides with the invariant subspace of the tensor product of the irreducible $SU(2)$ representations $[j_p]$ labeled by those spins and hence
\begin{eqnarray}
\label{proy}
\mathrm{dim}[\mathcal{H}^{\rm CS}(j_1,\ldots,j_n)]=\mathrm{dim}[\mathrm{Inv}(\otimes_p [j_p])]\,.
\end{eqnarray}
Here we will restrict ourselves to $|\gamma|\leq \sqrt{3}$. Therefore,  once the number $\mathrm{dim}[\mathrm{Inv}(\otimes_p [j_p])]$ is computed, the entropy can be directly obtained as its logarithm.

The problem of  determining the lists of spins satisfying a condition of the form $\sum_{p=1}^{n}\sqrt{j_p(j_p+1)}=a$ for a given value of $a$ has been already discussed in the literature \cite{prlnos}. In the following we use units such that $8\pi\gamma\ell_P^2=1$. In previous proposals an additional constraint, the so called projection constraint involving the sum of spin components $\sum_p m_p=0$, must be satisfied (see \cite{prlnos} and references therein for additional details). The role of the projection constraint is played now by the invariance condition  (\ref{proy}). In order to take it into account it is convenient to find first a suitable generating function giving this number for a given list of spins $j_1,\ldots, j_n$. In the following we will work with integer labels $k_p=2j_p$.  Since the result that we will find is closely related to the one corresponding to the Ghosh-Mitra (GM) counting \cite{GM}, we will carry the study of both proposals in parallel\footnote{Though the main results concerning the application of our methods to the GM counting have already appeared in the literature \cite{prlnos}, some new details are provided here for the first time.}. The dimension of the relevant invariant subspace can be derived from the scalar product of the characters $\chi_{k}$ of the representations $[k/2]$ of $SU(2)$ as
\begin{eqnarray*}
\mathrm{dim}[\mathrm{Inv}(\otimes_k [k/2]^{n_k})]&=&\langle\chi_0\,|\,\prod_{k} \chi^{n_k}_{k}\rangle_{SU(2)}=\frac{1}{\pi}\int_0^{2\pi} \mathrm{d}\theta \,\sin^2\theta \prod_{k} \left(\frac{\sin(k+1)\theta}{\sin\theta}\right)^{n_k}\\
&=&-\frac{1}{2\pi i}\oint_C \frac{\mathrm{d}z}{z} \frac{(z-z^{-1})^2}{2} \prod_{k} \left(\frac{z^{k+1}-z^{-k-1}}{z-z^{-1}}\right)^{n_k}\,,
\end{eqnarray*}
where we have considered the tensor product of $n_k$ representations of spin $k/2$ for the different values of $k$ considered. Here $C$ is the unit circle in the complex $z$-plane  defined by $z=e^{i\theta}$, $\theta\in[0,2\pi)$.

The previous formula should be compared with the one giving the number of solutions of the projection constraint for the GM counting that can be obtained in a similar fashion as the number of irreducible representations --taking into account multiplicities-- that appear in the direct sum decomposition of the tensor product $\otimes_p [k_p/2]$. This is given by
\begin{eqnarray*}
\left|\mathrm{rep}(\otimes_k [k/2]^{n_k}) \right|&=&
\sum_{l=0}^\infty\langle\chi_{l}\,|\,\prod_{k} \chi^{n_k}_{k}\rangle_{SU(2)}=\frac{1}{2\pi}\int_0^{2\pi} \mathrm{d}\theta \,\prod_{k} \left(\frac{\sin(k+1)\theta}{\sin\theta}\right)^{n_k}\\
&=&\frac{1}{2\pi i}\oint_C \frac{\mathrm{d}z}{z}  \prod_{k} \left(\frac{z^{k+1}-z^{-k-1}}{z-z^{-1}}\right)^{n_k}\,.
\end{eqnarray*}

The expressions given above allow us to identify the generating functions for the numbers that we want to obtain, namely
\begin{eqnarray*}
\mathrm{dim}[\mathrm{Inv}(\otimes_k [k/2]^{n_k})]&=&[z^0]\left(-\frac{(z-z^{-1})^2}{2} \prod_{k} \left(\frac{z^{k+1}-z^{-k-1}}{z-z^{-1}}\right)^{n_k}\right)\,,\\
\left|\mathrm{rep}(\otimes_k [k/2]^{n_k}) \right| &=&[z^0]\left(\prod_{k} \left(\frac{z^{k+1}-z^{-k-1}}{z-z^{-1}}\right)^{n_k}\right)\,,
\end{eqnarray*}
where $[z^0]F(z)$ denotes the coefficient of the $z^0$ term of the Laurent expansion of $F(z)$ around $z=0$. As in previous work, one has to take into account the relevant factors associated to the possible reordering of the $j$-labels in every `admissible' list of spins where, here,  admissible refers to the condition that they must satisfy an equality of the type $\sum_{p=1}^{n}\sqrt{j_p(j_p+1)}=a$. By proceeding as in \cite{EF} we get the following black hole generating functions
\begin{eqnarray}
G^{\rm ENP}(z;x_1,x_2,\cdots)&=&-\frac{(z-z^{-1})^2}{2}\left(1-\sum_{i=1}^\infty \sum_{\alpha=1}^\infty \Bigg( \frac{z^{k_\alpha^i+1}-z^{-k_\alpha^i-1}}{z-z^{-1}}\Bigg)\, x_i^{y_\alpha^i}\right)^{-1}\,,\label{GENP}\\
G^{\rm GM}(z;x_1,x_2,\cdots)&=&\left(1-\sum_{i=1}^\infty \sum_{\alpha=1}^\infty \Bigg( \frac{z^{k_\alpha^i+1}-z^{-k_\alpha^i-1}}{z-z^{-1}}\Bigg)\, x_i^{y_\alpha^i}\right)^{-1}\,.\label{GGM}
\end{eqnarray}
The integer numbers $k_\alpha^i$ and $y_\alpha^i$ appearing above are defined as follows:  For each squarefree positive integer $p_i$ the pairs  $(k_\alpha^i,y_\alpha^i)$, labeled by $\alpha\in\mathbb{N}$, are solutions to the Pell equation $(k+1)^2-p_i y^2=1$. In both cases, ENP and GM, $[z^0][x_1^{q_1}\cdots x_r^{q_r}]\,G(z;x_1,x_2,\cdots)$ provides the number of black hole states corresponding to a fixed area value $a$ such that $2a=q_1\sqrt{p_1}+\cdots+q_r\sqrt{p_r}$, where $q_i\in\mathbb{N}_0$ and $p_i$ are squarefree positive integers \cite{EF}.

The last step requires us to take into account the inequality appearing in the definition of the entropy (\ref{ineq}). The way to do this is to use Laplace-Fourier transforms as in \cite{EF1,M}. This is done by performing the substitution $x_i=e^{-s\sqrt{p_i}/2}$ and $z=e^{i\omega}$ in the previously obtained generating functions (\ref{GENP}) and (\ref{GGM}). By doing this we are left with complex functions on the variables $(s,\omega)\in \Omega\times [0,2\pi)$, where $\Omega\subset \mathbb{C}$  is certain region on the complex $s$-plane that can be easily determined. Explicitly,
\begin{eqnarray}
P^{\rm ENP}(s,\omega)&:=&G^{\rm ENP}(e^{i\omega};e^{-s\sqrt{p_1}/2},e^{-s\sqrt{p_2}/2},\cdots)\nonumber \\
&=&2\sin^2\omega\, \left(1- \displaystyle\sum_{k=1}^\infty \frac{\sin((k+1)\omega)}{\sin\omega}\,e^{-s\sqrt{k(k+2)}/2}\right)^{-1}\,,\label{Alejandro}
\\
P^{\rm GM}(s,\omega)&:=&G^{\rm GM}(e^{i\omega};e^{-s\sqrt{p_1}/2},e^{-s\sqrt{p_2}/2},\cdots)\nonumber \\
&=&  \left(1- \displaystyle\sum_{k=1}^\infty \frac{\sin((k+1)\omega)}{\sin\omega}\,e^{-s\sqrt{k(k+2)}/2}\right)^{-1}\label{GM}\,.
\end{eqnarray}
Notice that the only difference between (\ref{Alejandro}) and (\ref{GM}) is a  prefactor $2\sin^2\omega$.  The functions $P(s,\omega)$ defined above can be used to compute the entropy $S(a)$ of a black hole of area $a$ in the form
\begin{eqnarray}
S(a) = \log\left(\frac{1}{(2\pi)^2 i}\int_0^{2\pi}\mathrm{d}\omega\, \int_{x_0-i\infty}^{x_0+i\infty} \mathrm{d}s\, s^{-1} \, e^{as} P(s,\omega)\right)\,,\label{entropy}
\end{eqnarray}
where $x_0$ is any real number satisfying that, for all $\omega\in [0,2\pi)$, all the singularities of $f_\omega(s)=P(s,w)$ are confined to the band $\mathrm{Re}(s)<x_0$ of the complex $s$-plane. These formulas count the states corresponding to areas in the interval $(0,a]$ defined in (\ref{ineqDL}). From these one can immediately obtain the number of states in the interval $(a-\epsilon,a]$ and check that, asymptotically, both numbers grow in exactly the same way.

We end by discussing the asymptotic behavior of the entropy described by the formulas (\ref{Alejandro}), (\ref{GM}), and (\ref{entropy}). First notice that, by using arguments similar to the ones appearing in \cite{EF1}, one can show that for both (\ref{Alejandro}) and (\ref{GM}) the pole $\tilde{\gamma}>0$  responsible for the leading order in the asymptotic behavior (the one with the largest real part) is defined by
\begin{eqnarray}
\displaystyle\sum_{k=1}^\infty (k+1)\,e^{-\tilde{\gamma}\sqrt{k(k+2)}/2}=1\,.\label{gamma}
\end{eqnarray}
This means that the entropy in the new proposal indeed grows linearly with area and the value of the Immirzi parameter needed to reproduce the Bekenstein-Hawking law coincides with the one derived by Ghosh and Mitra $\gamma=\tilde{\gamma}/(2\pi)=0.274067\dots< \sqrt{3}$. On the other hand the logarithmic corrections for both models are different, in fact we get
$$
S^{\rm ENP}(a)=\frac{a}{4\ell_P^2}-\frac{3}{2}\log(a/\ell_P^2)+ O(1)\,,\quad S^{\rm GM}(a)=\frac{a}{4\ell_P^2}-\frac{1}{2}\log(a/\ell_P^2)+ O(1)\,.
$$
The reason behind this difference is the  prefactor $2\sin^2\omega$. In the vicinity of the largest real pole $\tilde{\gamma}$ (corresponding to $\omega=0$) defined by (\ref{gamma}) the poles of the integrand in the Laplace-Fourier transform given above can be approximated as
$$
\tilde{s}=\tilde{\gamma}-\tilde{\alpha}\omega^2+O(\omega^4)
$$
where $\tilde{\alpha}>0$ is a constant. In a neighborhood of $\omega=0$ we have $\sin^2\omega \sim \omega^2$ so the asymptotic behavior of the entropy for the new proposal \cite{ENP} is thus given by
\begin{eqnarray*}
S^{\rm ENP}(a)&\sim& \log\left(e^{\tilde{\gamma}a/(8\pi\gamma\ell_P^2)}\int_{-\varepsilon}^{\varepsilon}\mathrm{d}\omega\,\omega^2\exp (-\alpha  \omega^2 a/\ell^2_P) \right)\\
&\sim& \log\left(e^{\tilde{\gamma}a/(8\pi\gamma\ell_P^2)}\int_{-\infty}^{\infty}\mathrm{d}\omega\,\omega^2\exp (-\alpha  \omega^2 a/\ell^2_P) \right)\\
&\sim & \frac{a}{4\ell_P^2}-\frac{3}{2}\log (a/\ell_P^2)+O(1)\,,
\end{eqnarray*}
where $\alpha=\tilde{\alpha}/(8\pi\gamma)>0$.  In the Ghosh-Mitra case we have instead
\begin{eqnarray*}
S^{\rm GM}(a)&\sim& \log\left(e^{\tilde{\gamma}a/(8\pi\gamma\ell_P^2)}\int_{-\varepsilon}^{\varepsilon}\mathrm{d}\omega\,\exp (-\alpha  \omega^2a/\ell^2_P) \right)\\
&\sim& \log\left(e^{\tilde{\gamma}a/(8\pi\gamma\ell_P^2)}\int_{-\infty}^{\infty}\mathrm{d}\omega\,\exp (-\alpha  \omega^2 a/\ell^2_P)\right)\\
&\sim & \frac{a}{4\ell_P^2}-\frac{1}{2}\log(a/\ell^2_P)+O(1)\,.
\end{eqnarray*}
As we see the reason why the logarithmic correction in the Ghosh-Mitra case has a $-1/2$ coefficient is the absence of the $\omega^2$ factor in the previous integral.

A point of warning is needed here. As explained in \cite{EF1} the accumulation of the real parts of the poles of the integrands in (\ref{entropy}) may change the asymptotic behavior despite the fact that the exponential growth is well described by the pole with the largest real part. This means that the interesting substructure found in \cite{val1} may be present here too. In fact numerical computations for small black holes using both the brute force approach described in \cite{val1}, or a numerical implementation of the number theoretic methods encoded in the generating functions given above, show that the model considered here displays the same kind of interesting substructure in the entropy found in other instances. These computations also confirm the values of the Immirzi parameter and the $-3/2$ coefficient of the logarithmic term given by the new proposal of \cite{ENP}.

\bigskip

\noindent{\bf Acknowledgements.}
The authors want to thank Alejandro P\'erez for interesting discussions on this subject. The work was
partially supported by the Spanish MICINN research grants FIS2008-01980, FIS2008-03221, FIS2008-06078-C03-02 and ESP2007-66542-C04-01, and the Consolider-Ingenio 2010 Program CPAN (CSD2007-00042). I.A. has been partially supported by RGI founds of the Center of Gravitation and Cosmology at the University of Wisconsin-Milwaukee.


\begin{thebibliography}{99}

\expandafter\ifx\csname
natexlab\endcsname\relax\def\natexlab#1{#1}\fi
\expandafter\ifx\csname bibnamefont\endcsname\relax
\def\bibnamefont#1{#1}\fi
\expandafter\ifx\csname bibfnamefont\endcsname\relax
\def\bibfnamefont#1{#1}\fi
\expandafter\ifx\csname citenamefont\endcsname\relax
\def\citenamefont#1{#1}\fi
\expandafter\ifx\csname url\endcsname\relax
\def\url#1{\texttt{#1}}\fi
\expandafter\ifx\csname urlprefix\endcsname\relax\def\urlprefix{URL
}\fi \providecommand{\bibinfo}[2]{#2}
\providecommand{\eprint}[2][]{\url{#2}}


\bibitem{ENP}
\bibinfo{author}{\bibfnamefont{J.}\bibnamefont{~Engle}},
\bibinfo{author}{\bibfnamefont{K.}\bibnamefont{~Noui}},
\bibnamefont{and}
\bibinfo{author}{\bibfnamefont{A.}\bibnamefont{~Perez}},
\textit{Black hole entropy and SU(2) Chern-Simons theory},
\eprint{arXiv:0905.3168}.




\bibitem{ABCK}
\bibinfo{author}{\bibfnamefont{A.}\bibnamefont{~Ashtekar}},
\bibinfo{author}{\bibfnamefont{J.}\bibnamefont{~Baez}},
\bibinfo{author}{\bibfnamefont{A.}\bibnamefont{~Corichi}},
\bibnamefont{and}
\bibinfo{author}{\bibfnamefont{K.}\bibnamefont{~Krasnov}},
\bibinfo{journal}{Phys. Rev. Lett.} \textbf{\bibinfo{volume}{80}},
\bibinfo{pages}{904} (\bibinfo{year}{1998}).

\bibitem{ABK}
\bibinfo{author}{\bibfnamefont{A.}\bibnamefont{~Ashtekar}},
\bibinfo{author}{\bibfnamefont{J.}\bibnamefont{~Baez}},
\bibnamefont{and}
\bibinfo{author}{\bibfnamefont{K.}\bibnamefont{~Krasnov}},
\bibinfo{journal}{Adv. Theor. Math. Phys. } \textbf{\bibinfo{volume}{4}},
\bibinfo{pages}{1} (\bibinfo{year}{2000}).


\bibitem{DL}
\bibinfo{author}{\bibfnamefont{M.}\bibnamefont{~Domagala}}
\bibnamefont{and}
\bibinfo{author}{\bibfnamefont{J.}\bibnamefont{~Lewandowski}},
\bibinfo{journal}{Class. Quant. Grav.} \textbf{\bibinfo{volume}{21}},
\bibinfo{pages}{5233} (\bibinfo{year}{2004}).

\bibitem{GM}
\bibinfo{author}{\bibfnamefont{A.}\bibnamefont{~Ghosh}}
\bibnamefont{and}
\bibinfo{author}{\bibfnamefont{P.}\bibnamefont{~Mitra}},
\bibinfo{journal}{Phys. Lett. B\textbf{616}, 114 (2005)}.


\bibitem{prlnos}
\bibinfo{author}{\bibfnamefont{I.}\bibnamefont{~Agull\'o}},
\bibinfo{author}{\bibfnamefont{J.~F.}\bibnamefont{~Barbero G.}},
\bibinfo{author}{\bibfnamefont{E.}\bibnamefont{~Fernandez-Borja}},
\bibinfo{author}{\bibfnamefont{J.}\bibnamefont{~D\'{\i}az-Polo}},
\bibnamefont{and}
\bibinfo{author}{\bibfnamefont{E.~J.~S.}\bibnamefont{~Villase\~nor}},
\bibinfo{journal}{Phys. Rev. Lett. \textbf{100},  211301 (2008)}.

\bibitem{EF}
\bibinfo{author}{\bibfnamefont{J.~F.}\bibnamefont{~Barbero G.}}
\bibnamefont{and}
\bibinfo{author}{\bibfnamefont{E.~J.~S.}\bibnamefont{~Villase\~nor}},
\bibinfo{journal}{Phys. Rev. D\textbf{77}, 121502(R) (2008)}.

\bibitem{EF1}
\bibinfo{author}{\bibfnamefont{J. F.}\bibnamefont{~Barbero G.}}
\bibnamefont{and}
\bibinfo{author}{\bibfnamefont{E.~J.~S.}\bibnamefont{~Villase\~nor}},
\bibinfo{journal}{Class. Quant. Grav.} \textbf{\bibinfo{volume}{26}},
\bibinfo{pages}{035017} (\bibinfo{year}{2009}).


\bibitem{val}
\bibinfo{author}{\bibfnamefont{I.}\bibnamefont{~Agullo}},
\bibinfo{author}{\bibfnamefont{E.~F.}\bibnamefont{~Borja}},
\bibnamefont{and}
\bibinfo{author}{\bibfnamefont{J.}\bibnamefont{~Diaz-Polo}},
\bibinfo{journal}{JCAP\textbf{07}, 016 (2009)}.


\bibitem{KM}
\bibinfo{author}{\bibfnamefont{R.~K.}\bibnamefont{~Kaul}}
\bibnamefont{and}
\bibinfo{author}{\bibfnamefont{P.}\bibnamefont{~Majumdar}},
\bibinfo{journal}{Phys. Lett. B\textbf{439}, 267 (1998)}.


\bibitem{M}
\bibinfo{author}{\bibfnamefont{K.}\bibnamefont{~Meissner}},
\bibinfo{journal}{Class. Quant. Grav.} \textbf{\bibinfo{volume}{21}},
\bibinfo{pages}{5245} (\bibinfo{year}{2004}).

\bibitem{val1}
\bibinfo{author}{\bibfnamefont{A.}\bibnamefont{~Corichi}},
\bibinfo{author}{\bibfnamefont{J.}\bibnamefont{~Diaz-Polo}},
\bibnamefont{and}
\bibinfo{author}{\bibfnamefont{E.}\bibnamefont{~Fernandez-Borja}},
\bibinfo{journal}{Phys. Rev. Lett.} \textbf{\bibinfo{volume}{98}},
\bibinfo{pages}{181301} (\bibinfo{year}{2007}).
\bibinfo{author}{\bibfnamefont{A.}\bibnamefont{~Corichi}},
\bibinfo{author}{\bibfnamefont{J.}\bibnamefont{~Diaz-Polo}},
\bibnamefont{and}
\bibinfo{author}{\bibfnamefont{E.}\bibnamefont{~Fernandez-Borja}},
\bibinfo{journal}{Class. Quant. Grav.} \textbf{\bibinfo{volume}{24}},
\bibinfo{pages}{243} (\bibinfo{year}{2007}).








\end{thebibliography}
\end{document}